\documentstyle[12pt,psfig]{article}
\newcommand{\nsect}{\setcounter{equation}{0}
\def\theequation{\thesection.\arabic{equation}}\section}

\catcode`\@=11
\def\marginnote#1{}
\def\ifmath#1{\relax\ifmmode #1\else $#1$\fi}

\def\bold#1{\setbox0=\hbox{$#1$}%
     \kern-.025em\copy0\kern-\wd0
     \kern.05em\copy0\kern-\wd0
     \kern-.025em\raise.0433em\box0 }

\def\GENITEM#1;#2{\par\vskip6pt \hangafter=0 \hangindent=#1
   \Textindent{$ #2$ }\ignorespaces}

\def\Vbu{V_B^{(1)}}
\def\Vbd{V_B^{(2)}}
\def\Vfu{V_F^{(1)}}

 \def\Vbf{V_{BF}^{(2)}}
\newcount\hour
\newcount\minute
\newtoks\amorpm
\hour=\time\divide\hour by60
\minute=\time{\multiply\hour by60 \global\advance\minute by-
\hour}
\edef\standardtime{{\ifnum\hour<12 \global\amorpm={am}%
    \else\global\amorpm={pm}\advance\hour by-12 \fi
    \ifnum\hour=0 \hour=12 \fi
    \number\hour:\ifnum\minute<100\fi\number\minute\the\amorpm}}
\edef\militarytime{\number\hour:\ifnum\minute<100\fi\number\minute}
\def\draftlabel#1{{\@bsphack\if@filesw {\let\thepage\relax
  \xdef\@gtempa{\write\@auxout{\string
    \newlabel{#1}{{\@currentlabel}{\thepage}}}}}\@gtempa
    \if@nobreak \ifvmode\nobreak\fi\fi\fi\@esphack}
     \gdef\@eqnlabel{#1}}
\def\@eqnlabel{}
\def\@vacuum{}
\def\draftmarginnote#1{\marginpar{\raggedright\scriptsize\tt#1}}
\def\draft{\oddsidemargin -.5truein
        \def\@oddfoot{\sl preliminary draft \hfil
        \rm\thepage\hfil\sl\today\quad\militarytime}
        \let\@evenfoot\@oddfoot \overfullrule 3pt
        \let\label=\draftlabel
        \let\marginnote=\draftmarginnote

\def\@eqnnum{(\theequation)\rlap{\kern\marginparsep\tt\@eqnlabel}%
\global\let\@eqnlabel\@vacuum}  }
\def\preprint{\twocolumn\sloppy\flushbottom\parindent 1em
        \leftmargini 2em\leftmarginv .5em\leftmarginvi .5em
        \oddsidemargin -.5in    \evensidemargin -.5in
        \columnsep 15mm \footheight 0pt
        \textwidth 250mmin      \topmargin  -.4in
        \headheight 12pt \topskip .4in
        \textheight 175mm
        \footskip 0pt

\def\@oddhead{\thepage\hfil\addtocounter{page}{1}\thepage}
        \let\@evenhead\@oddhead \def\@oddfoot{} \def\@evenfoot{}
}
\def\titlepage{\@restonecolfalse\if@twocolumn\@restonecoltrue\onecolumn
     \else \newpage \fi \thispagestyle{empty}\c@page\z@
        \def\thefootnote{\fnsymbol{footnote}} }
\def\endtitlepage{\if@restonecol\twocolumn \else  \fi
        \def\thefootnote{\arabic{footnote}}
        \setcounter{footnote}{0}}  
\catcode`@=12
\relax

\def\be{\begin{equation}}
\def\ee{\end{equation}}
\def\bea{\begin{eqnarray}}
\def\eea{\end{eqnarray}}

\def\NPB#1#2#3{{\it Nucl.~Phys.} {\bf{B#1}} (19#2) #3}
\def\PLB#1#2#3{{\it Phys.~Lett.} {\bf{B#1}} (19#2) #3}
\def\PRD#1#2#3{{\it Phys.~Rev.} {\bf{D#1}} (19#2) #3}

\def\PTP#1#2#3{{\it Prog.~Theor.~Phys.} {\bf#1}  (19#2) #3}

\def\mst1{m_{\;\widetilde{t}_{1}}}

\def\st{\;\widetilde{t}}

\def\mst2{m_{\;\widetilde{t}_{2}}}
\def\mst12{m_{\;\widetilde{t}_{1,2}}}

\def\MSbar{\overline{\rm MS}}
\def\DRbar{\overline{\rm DR}}
\def\msb1{m_{\;\widetilde{b}_{1}}}
\def\msb2{m_{\;\widetilde{b}_{2}}}
\def\msb12{m_{\;\widetilde{b}_{1,2}}}

\def\mtilde2{\widetilde{m}^{2}}

\relax

\begin{document}
\topmargin-2.5cm

%
\begin{titlepage}
\begin{flushright}
IEM-FT-181/98 \\
hep--ph/9809275 \\
\end{flushright}
\vskip 0.3in
\begin{center}{\Large\bf The effective
potential in the presence of several 
mass scales\footnote{Work supported in part by the European Union
(contract CHRX/CT92-0004) and CICYT of Spain
(contract AEN95-0195).} }
\vskip .5in
{\bf J.A. Casas, V. Di Clemente and  M. Quir{\'o}s} \\
\vskip.35in
Instituto de Estructura de la Materia, CSIC, Serrano
123, 28006-Madrid, Spain
\end{center}
\vskip2.cm
\begin{center}
{\bf Abstract}
\end{center}
\begin{quote}
We consider the problem of improving the effective potential in
mass independent schemes, as e.g. the $\MSbar$ or $\DRbar$
renormalization scheme, in the presence
of an arbitrary number of fields with $\phi$-dependent
masses $M_i(\phi_c)$. We use the
decoupling theorem at the scales $\mu_i=M_i(\phi_c)$ such that the
matching between the effective (low energy) and complete (high
energy) one-loop theories contains no thresholds. 
We find that for any value of $\phi_c$,
there is a convenient scale $\mu^*\equiv\min_i\{M_i(\phi_c)\}$, 
at which the loop
expansion has the best behaviour and the effective potential has the least
$\mu$-dependence. Furthermore, at this scale the
effective potential coincides with the (improved) tree-level one
in the effective field theory. 
The decoupling method is explicitly illustrated with a simple Higgs-Yukawa
model, along with its relationship with other decoupling
prescriptions and with proposed multi-scale renormalization 
approaches. The procedure leads to a nice suppression of potentially 
large logarithms and can be easily adapted to include higher-loop effects,
which is explicitly shown at the two-loop level.
\end{quote}

\vskip1.cm

\begin{flushleft}
IEM-FT-181/98\\
September 1998 \\
\end{flushleft}
\end{titlepage}
\setcounter{footnote}{0}
\setcounter{page}{0}
\newpage

\nsect{Introduction}

Unlike the case of the standard model (SM), where one-loop radiative
corrections to the effective potential are essentially dominated
by the top quark, all extensions thereof involve new mass scales 
corresponding to new physics. A well celebrated example is the
supersymmetric standard model (MSSM)~\cite{HK} where, beside the top quark
scale $m_t$, there are mass scales corresponding to the stops
$m_{\st_{L,R}}$ which are strongly coupled to the Higgs
sector and thus provide very important contributions to the
effective potential: it is a typical multi-scale case.
Moreover, if there is a large mixing in the stop sector, the two stop
mass eigenvalues become very split, which represents by itself a
multi-scale problem.
Understanding the behaviour of the effective potential in
multi-scale cases is then essential to handle a number of
physical problems associated with them. One example is
the supersymmetric standard model and, in particular, the
calculation of the Higgs boson mass spectrum from the effective
potential~\cite{higgsmass}. A good treatment of the effective potential would
permit, in particular, to attack the above-mentioned case of large
stop splitting, which still remains as an open question.
Another application would be to establish the relationship between the
scale of new physics and the possible instability scale of the
standard model~\cite{instability}. This precise relation can be of the highest
interest specially in the case that the standard model Higgs
boson is light. 

Let us briefly review the state of the art concerning the computation of
the effective potential, focussing in the case of multi-scale
scenarios. The improved effective potential in any mass-independent
renormalization scale can be expanded in a loop expansion as~\cite{CW}:
\be
V_{\rm eff}(\lambda_a(t),\phi(t),\mu(t))=V^{(0)}+V^{(1)}+\cdots
\label{expansion}
\ee
where $\lambda_a(t)$ are all dimensionless and dimensionful
couplings of the theory, running with the 
renormalization group equation (RGE) scale
$\mu(t)\equiv e^t\mu$, and $\phi(t)=\xi(t)\phi_c$ is (are) the running
field(s) with $\xi(t)=\exp(-\int_0^t \gamma(t')dt')$, $\gamma(t)$
being the anomalous dimension of the $\phi$-field. In the loop
expansion (\ref{expansion}), $V^{(0)}$ is the improved tree
level potential while the one-loop correction can be written as
\bea
V^{(1)}&=&\kappa \sum_i V_i^{(1)}\nonumber \\
V_i^{(1)}&=&\frac{n_i}{4} M_i^4(\phi)\log\frac{M_i^2(\phi)}
{\mu_i^2}
\label{unloop}
\eea
where $\kappa=1/16\pi^2$,
$n_i$ is the number of degrees of freedom of the $i$-th
field (negative for fermions), $M_i^2(\phi)$ are the (tree-level)
mass eigenvalues,
$\mu_i^2\equiv e^{C_i}\bar\mu^2$ ($\bar\mu$ is the
renormalization scale in the corresponding scheme, typically
$\MSbar$ or $\DRbar$), and $C_i$ are scheme dependent
constants~\footnote{For instance, in the $\MSbar$ scheme, $C_i$=
3/2 (5/6) for scalar bosons and fermions (gauge bosons), while
$C_i$=3/2 for all fields in the $\DRbar$ scheme. In a theory
where all fields have equal value of $C_i$, as e.g. in a theory
without gauge bosons, renormalized in the $\MSbar$ scheme, or an
arbitrary theory renormalized in the $\DRbar$ scheme, we can
reabsorb all finite constants in a redefinition of the
renormalization scale: $e^{C_i}\mu^2\rightarrow\mu^2$. This will
be the case in the simple example considered in this paper.}.   

It is evident from (\ref{unloop}) that in the case of 
a {\em single} mass scale, say $M_i(\phi)= m(\phi)$, a
judicious choice of the renormalization scale $\mu^2\sim m^2(\phi)$
removes large logarithms from the one-loop corrections.
Actually, this choice amounts to a resummation of all the
leading-logarithms at all order in loop perturbation 
theory~\cite{CW} (if the
$\lambda_a, \phi$ parameters are running with the two-loop RGE, the
next-to-leading logarithms are also resummed~\cite{todos}). However,
if there are very different mass scales, e.g. $m(\phi)$ and $M(\phi)$,
any choice of the renormalization scale will leave large
logarithms in (\ref{unloop}), as $\log[M^2(\phi)/m^2(\phi)]$ which
jeopardize the validity of perturbation theory.

There are two proposed solutions in the literature to the
multi-scale improved effective potential. One, introduced by
Einhorn and Jones~\cite{EJ} , and recently modified by Ford and
Wiesendanger~\cite{FW1}, proposing the presence of {\it independent}
 renormalization
scales $\mu_i$ which can then be adjusted differently to take
care of the different mass scales. This procedure is however
highly heterodox, involves as many RGE for each parameter 
as mass scales in the
problem and seems very difficult to be applied to realistic
cases with possibly many mass scales. Another solution, more
physical in our opinion, was studied by Bando, Kugo, Maekawa and
Nakano~\cite{BKMN}, and consists in recognizing that the origin
of the problem in mass-independent renormalization
schemes is that the decoupling is not automatically included in the formalism and
has to be incorporated~\footnote{A modified mass dependent renormalization
scheme has been proposed in Ref.~\cite{massdep} to handle with the multimass problem.}. 
Then all masses smaller than a given
scale behave as massless while larger masses decouple and never
generate problems with perturbation theory. Below a decoupling
scale, the theory is an effective field theory with new RGE and,
eventually, with threshold effects taking care of the matching
of both theories at the decoupling scale~\cite{georgi}.
These authors, however, have
identified the decoupling scale of the $i$-th particle
with the $\phi$-independent mass $m_i$ eventually contributing to the
mass eigenvalue $M_i(\phi)$. This is not a very convenient
and/or physical choice because: {\bf a)} In the presence of the
background field $\phi$, particles propagate with the mass
$M_i(\phi)$, which is therefore the relevant scale for decoupling;
{\bf b)} For large values of the field $\phi$, such that the main
contribution to the mass $M_i(\phi)$ does not come from $m_i$,
it cannot be a good approximation; {\bf c)} It does not
contemplate the case where the decoupled particle does not have
an invariant mass; {\bf d)} It does not fit some
generic situations, as e.g. the case where there are large
mixings between different states. This is the case of the stops
$\st_{L,R}$ where besides the invariant masses $m_{\st_L}$ and
$m_{\st_R}$ there is the mixing parameter $A_t$. In that case
decoupling at $m_{\st_L}$ and $m_{\st_R}$ would be clearly inappropriate
for large mixing. In addition, the choice of the value of the
renormalization scale proposed in Ref.~\cite{BKMN} in order to evaluate
the potential, namely $\mu$ equal to the fermion mass, is only
well defined when there is a single fermion in the model and, even in
that case, it turns out to be inappropriate in many instances, as we
will see.

In this paper we will consider the effective potential in the
multi-scale case, proposing a new and simple approach based upon the
decoupling theorem~\cite{AC}. The method is fully consistent with perturbation theory, 
even when there are very different scales in the game, which normally 
leads to potentially dangerous logarithms.
We will
illustrate the method with a simple Higgs-Yukawa model and
present the rationale to be applied to more realistic and
physical models. The general decoupling method is introduced in
Section 2, and applied to a simple Higgs-Yukawa model in Section
3, where the numerical analysis has been explicitly performed
(in the one-loop approximation) 
and illustrated with plots. In Section 4 we compare our
decoupling approach with the multi-scale renormalization
proposals. In Section 5 we
generalize the analysis of Section 3 by including two-loop
corrections. We explicitly show how the dangerous logarithms continue
to be suppressed at this level, confirming the perturbative validity
of the approach. Section 6 is devoted to our conclusions and outline.

\nsect{Our method}

Our proposal in this paper is to use the decoupling theorem~\cite{AC} to
handle with the problem of multi-scale in the effective
potential, using as decoupling scale for the mass
$M_i(\phi)$ the scale
\be
\mu_i^d=M_i(\phi)
\label{escalai}
\ee
In other words we replace the improved potential (\ref{unloop}) by
\be
V^{(1)}=\kappa\sum_i V_i^{(1)}\;\theta_i
\label{unloopdec}
\ee
where $\theta_i\equiv \theta(\mu_i-M_i(\phi))$ is the Heaviside
theta function. Using the above prescription we obtain, as a
bonus, that there are {\em no} one-loop threshold corrections neither 
for the couplings nor for the effective potential. In this way the
effective theory is the usual one, containing just the light degrees
of freedom while all the effects of integrating out the heavy
modes are encoded in the initial conditions of the low energy
couplings at the decoupling scale $\mu_i^d$. Let us recall that
in a mass-independent renormalization scheme, such as 
the $\MSbar$ or $\DRbar$, the decoupling must be implemented by hand,
being justified by physical reasons~\footnote{In section
3 we will discuss how to incorporate the decoupling in this kind of
schemes in a natural way, showing that
the previous procedure is favoured not just by physical considerations
but also by the demand of highest calculational accuracy.}.

Since the effective potential satisfies the RGE
\be
{\cal D}V\equiv
\left(\mu\frac{\partial}{\partial\mu}+\beta_a\frac{\partial}{\partial\lambda_a}
-\gamma\phi\frac{\partial}{\partial\phi}\right)V=0
\label{RGEpot}
\ee
the decoupling prescription for the $\beta$ functions is
inherited from the corresponding one imposed on the effective
potential. For instance, at one-loop:
\be
\left(\beta_a\frac{\partial}{\partial\lambda_a}
-\gamma\phi\frac{\partial}{\partial\phi}\right)V^{(0)}= 
\frac{1}{32\pi^2}\sum_i n_i M_i^4\theta_i
\label{RGEgen}
\ee
and so the $\beta$ and $\gamma$-functions decouple at the same points as the
effective potential (\ref{unloopdec}). Notice that, as for the effective
potential, with this prescription there are no one--loop threshold
corrections for the $\lambda_a$ couplings~\footnote{Observe that
$\theta_i\equiv\theta\left(-V_i^{(1)}\right)$ and, as a consequence,
$V_i^{(1)}{\cal D}\theta_i=-V_i^{(1)}\delta\left(V_i^{(1)}\right)
{\cal D}V_i^{(1)}\equiv 0$. Therefore the presence of Heaviside
$\theta$-functions in the effective potential does not yield
extra $\delta$-function terms to the RGE which would translate,
upon integration, into threshold effects for the coupling
constants.}. For a fixed value of the
classical field, $\phi_c$, the effective potential given by
Eqs.~(\ref{expansion}), (\ref{unloopdec}), with the $\lambda_a$ parameters
running in this way, represents the one--loop approximation to $V(\phi_c)$,
which is $\mu$--independent (see Eq.~(\ref{RGEpot})) up to two loop
corrections. These corrections, however, may be very large due to the
presence of potentially large logarithms in $V^{(n)}$. Thus, one has to
be careful in choosing a value of $\mu$ which gives the finest
approximation to $V(\phi_c)$. We will come back to this point shortly.

Let us now turn to a subtle point. It is usually assumed that the
initial conditions for the $\lambda_a(\mu)$ parameters are given at some
large value of the scale (larger than all the involved masses), which can
be even at some grand unified value. This is reasonable since the SM is
supposed to be a low-energy effective theory coming from a more
fundamental theory. The values of $\lambda_a(\mu)$ at low scales
are the result of the RGE running from the high-energy scale,
i.e. a top--down procedure.
In between both scales, the successive particles will be decoupled
according to the values given by (\ref{escalai}). Since the decoupling
scales $\mu_i^d$ depend on $\phi_c$, it is clear that the 
low-energy values of $\lambda_a(\mu)$ are (logarithmically)
$\phi_c$--dependent, so we can
denote them as $\lambda_a(\mu,\phi_c)$. Of course,
in many physical instances what is known is the ``experimental'' values of
the parameters, $\lambda_a^{\rm exp}$. These correspond
to the values of $\lambda_a(\mu=\mu_0, \phi_c=v)$, where $\mu_0$
is the low-energy scale at which
experiments are done and $v=\langle \phi_c\rangle$ is
the {\em physical} VEV of the scalar field~\footnote{From a low-energy
point of view, what actually makes sense are the coefficients of the 
$n$-point operators $(\phi-v)^n$. These can be straightforwardly  related 
to the $\lambda_a(\mu=\mu_0, \phi_c=v)$ couplings.
Notice that, in a precise way, the (logarithmic) dependence of
$\lambda_a$ on $\phi$ should be taken into account when 
establishing this connection.}. In that case, the RGE
running and thresholds (\ref{escalai}) permit to evaluate in a bottom--up
procedure the couplings $\lambda_a(\mu,v)$ at any large scale. Notice
that for a large enough value of $\mu$, say $\mu$
greater than all masses $M_i(\phi_c)$, the values
of $\lambda_a(\mu,\phi_c)$ do not depend on $\phi_c$. Then, one can
come back downstairs (top--down procedure) with the RGE and any desired
value of $\phi_c$, using the corresponding decouplings. In this way one
can reconstruct the $\phi_c$ dependence of $\lambda_a$ for any scale
$\mu$. This will be illustrated in the next section.

Let us now examine what is the optimum value of the renormalization
scale, say $\mu=\mu^*$, at which the one-loop potential $V_{eff}(\phi_c,
\mu)$ should be evaluated. There are two sensible criteria for this
choice. The first one is to take $\mu^*$ as the scale at which the
one-loop potential has the least $\mu$--dependence. This is reasonable since
the {\em complete} potential satisfies $dV/d\mu\equiv0$ and a
substantial dependence of the one--loop potential on $\mu$ clearly
induces large uncertainties in the evaluation itself. The second criterion is
to take $\mu^*$ as the scale at which the loop expansion has the
best apparent behavior, i.e. $V^{(1)}=0$. Again, a substantial departure from
this criterion amounts to large uncertainties concerning higher order
corrections. Obviously, the most desirable situation occurs when the two
criteria give (essentially) the same value of $\mu^*$.

It is clear that for scales below $\min_i\{M_i(\phi)\}$
the effective potential is contributed only by the tree
level potential. Hence any $\mu^*\leq \min_i\{M_i(\phi)\}$ 
trivially satisfies the
second criterion, $V^{(1)}=0$, as well as the first criterion, 
$dV_{eff}/d\mu\equiv0$, since all the particles get decoupled from the
RGE. Consequently, this will be our criterion to choose $\mu^*$.
Nevertheless, it is not obvious that, in particular models, 
there are not other values of $\mu$ satisfying the required conditions.
This will be examined in the next section.

Including two-loop corrections to the previous approach is
straightforward. It can be done in two possible ways: Either by modifying the
prescription (\ref{escalai}) for the decoupling scales $\mu_i^d$ in order
to cancel one--
plus two--loop corrections to the effective potential for every particle
species, or just maintaining prescription (\ref{escalai}) in which case
there will appear two-loop threshold effects in the potential and in
the couplings at every decoupling scale. 

In the next section we will illustrate numerically the above ideas with
a simple Higgs-Yukawa model and postpone more realistic theories, as
extensions of the standard model, for a future analysis although it is
clear from the present section that the generalization is straightforward.

\nsect{The Higgs-Yukawa model}

This model contains a Higgs boson $\phi$ and an arbitrary number $N$ of Dirac
fermions $\psi_i$ coupled to the Higgs with a Yukawa coupling $g$. 
The lagrangian is given by
\bea
{\cal L}=\frac{1}{2}(\partial\phi)^2 - \frac{1}{2}m^2\phi^2 -
\frac{1}{4!}\lambda\phi^4 + \bar\psi(i\slash\!\!\!\partial + g\phi)\psi
- \Lambda\ ,
\label{lagr}
\eea
where $\psi\equiv(\psi_1,...\psi_N)$ and $\Lambda$ is a cosmological
constant term. Using the
decoupling prescription proposed in (\ref{unloopdec}), the 
improved one-loop effective
potential and $\beta$ and $\gamma$-functions for this model can be written as: 
\bea
V^{(0)}&=&\frac{1}{2}m^2\phi^2+\frac{1}{4!}\lambda\phi^4+\Lambda
\nonumber\\
V^{(1)}&=&\kappa\left( V_B^{(1)}\theta_B+V_F^{(1)}\theta_F\right)\nonumber\\
V_B^{(1)}&=& \frac{1}{4}\;M_B^4\log\frac{M_B^2}{\mu^2}\nonumber\\ 
V_F^{(1)}&=& - N M_F^4\;\log\frac{M_F^2}{\mu^2}
\label{unloopmod}
\eea
where 
\bea
M_B^2=\frac{1}{2}\lambda\phi^2+m^2,\;\; M_F^2=g^2\phi^2
\label{MfMb}
\eea
are the Higgs
boson and fermion masses, respectively,
$\theta_{B,F}=\theta(\mu-M_{B,F})$, $\mu^2\equiv\bar\mu^2 e^{3/2}$, and 
\bea
\beta_\lambda&=&\frac{1}{(4\pi)^2}\left[ 3\lambda^2 \theta_B+ 8 g^2 N
\left(\lambda-6 g^2 \right)\theta_F \right] 
\nonumber\\
\beta_{m^2}&=&\frac{1}{(4\pi)^2}\left[\lambda \theta_B+ 4 N g^2 \theta_F
\right]
m^2 \nonumber \\
\beta_\Lambda&=&\frac{1}{(4\pi)^2} \frac{1}{2}m^4 \theta_B 
\nonumber \\
\beta_{g^2}&=&\frac{1}{(4\pi)^2} \left[ 6\theta_B\theta_F+4 N \theta_F\right]
g^4 
\nonumber \\
&=&\frac{1}{(4\pi)^2} \left[ 6\theta(M_B-M_F)\theta_B+
\left(6\theta(M_F-M_B)+4N\right) \theta_F\right] g^4 
\nonumber \\
\gamma_\phi&=&\frac{1}{(4\pi)^2} 2 g^2 N \theta_F
\label{RGE}
\eea

\begin{figure}[h]
\mbox{
\psfig{figure=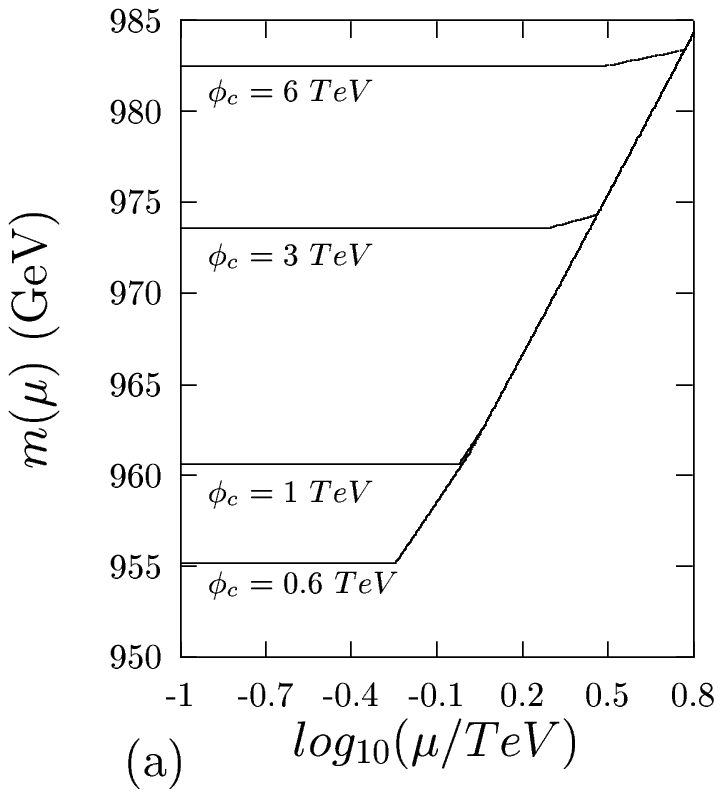,bbllx=5.3cm,bblly=16cm,bburx=13cm,bbury=23.5cm,height=5.8cm,width=6.5cm} \hspace{2cm}  \psfig{figure=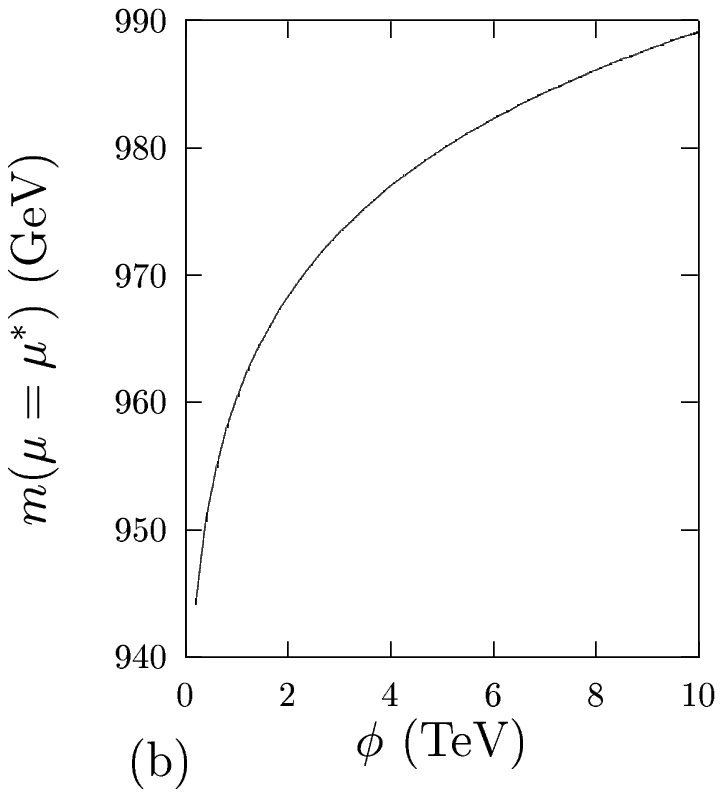,bbllx=5.3cm,bblly=16cm,bburx=13cm,bbury=23.5cm,height=5.8cm,width=6.5cm}}
\caption{\small{a) Plot of the $m$ parameter as a function of $\mu$ for several values of $\phi_c$. b) Plot of the $m(\mu^*)$ parameter as a function of $\phi$, where $ \mu^* = \min\{M_F,M_B\}$. In both figures, $\lambda=0.1$, $g=1$ and $m=1$ TeV at high scale $\mu_M \simeq 2\times 10^{4}$ GeV.}}
\label{fig:masa}
\end{figure}

\begin{figure}[h]
\vspace{-3.3cm}
\mbox{
\psfig{figure=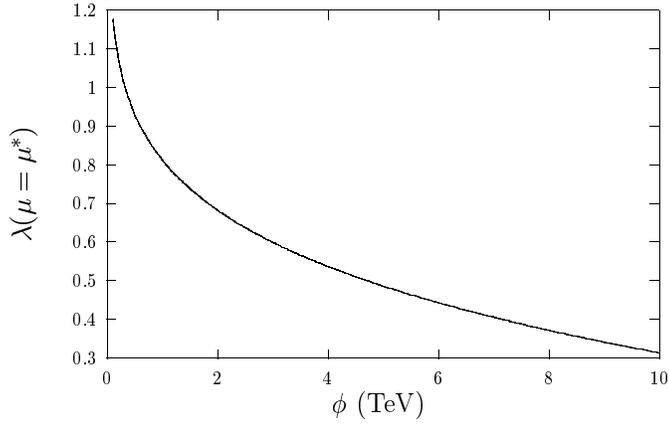,height=20cm}}
\vspace{-12cm}
\caption{\small{ Plot of the $\lambda(\mu^*)$ parameter as a function of $\phi$, where again $ \mu^* = \min\{M_F,M_B\}$ and the initial condition are the same as in Fig.~\ref{fig:masa}.}}
\label{fig:lambda}
\end{figure}

All the couplings, including the mass parameter, and the field $\phi$ in
(\ref{unloopmod}) are running with the renormalization scale $\mu$.  
As described in the previous section, we will start with initial
boundary conditions for all the couplings at a high scale. In 
particular we will fix $\lambda=0.1$, $g=1$ and $m=1$ TeV at the high scale
$\mu_M=e^{3/4}10^{4}$ GeV $\simeq 2\times 10^{4}$ GeV. 
Then, running the RGE in a top--down procedure,
passing through the various mass thresholds of the model, we can get the 
couplings $\lambda,m$, etc. at any value of $\mu$ 
(we will run downwards until the low scale $\mu_m=100$ GeV).
As stressed in the previous section, since the decoupling scales
associated with the thresholds (in
this case $M_B$ and $M_F$) are $\phi_c$-dependent, so the    
low-energy parameters are. 
This is illustrated in Fig.~\ref{fig:masa}a
for the $m$ parameter and several values of $\phi_c$. 
There is a value of $\mu$, corresponding to 
$\max\{M_F,M_B\}$, at which the heaviest particle decouples and the
top--down running of $m(\mu)$ departures from the complete RGE. 
Then, after passing
through the lower threshold, at $\min\{M_F,M_B\}$, the value of
$m(\mu)$ becomes frozen. Since both $M_B$ and $M_F$ are 
$\phi_c$-dependent, see (\ref{MfMb}), so the low-energy value of $m(\mu)$
is. An analogous behavior occurs for the remaining parameters.
Figs.~\ref{fig:masa}b,~\ref{fig:lambda} show explicitly the variation of
$m(\mu_m)$ and $\lambda(\mu_m)$ with $\phi_c$.

\begin{figure}[h]
\mbox{
\psfig{figure=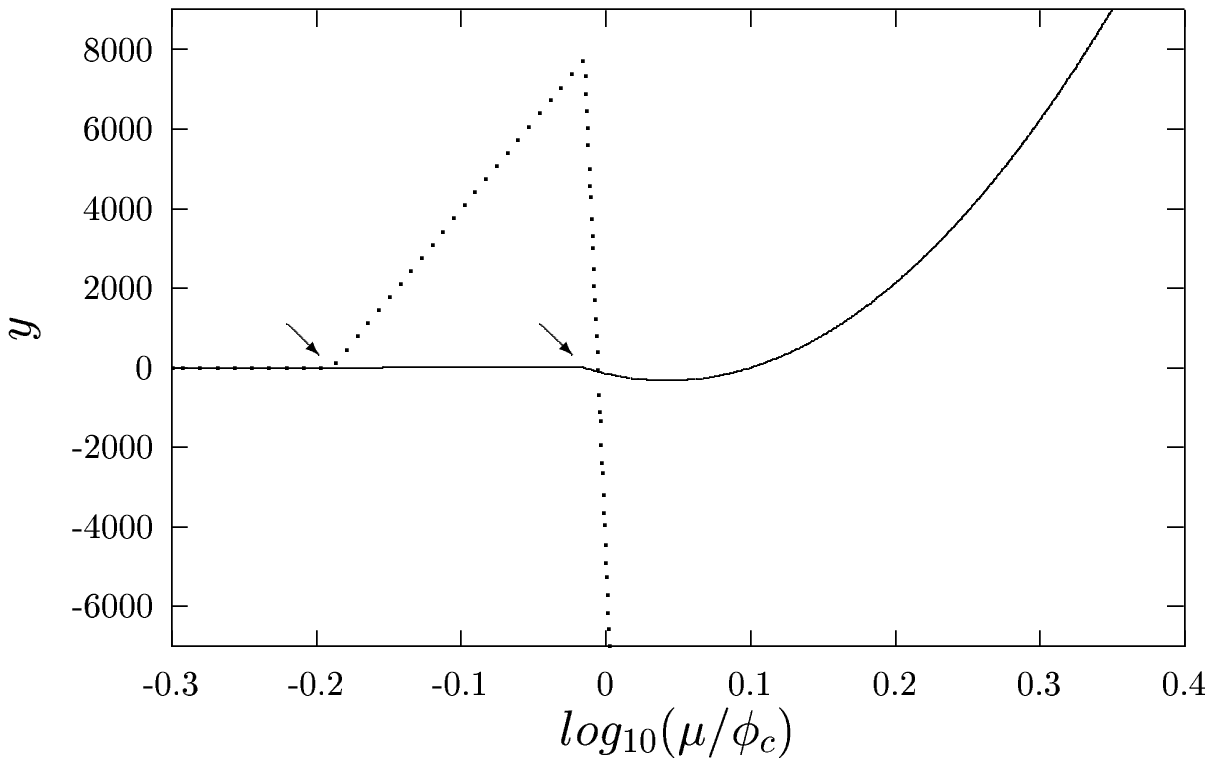,bbllx=5.6cm,bblly=16cm,bburx=13cm,bbury=24cm,height=6cm,width=4.5cm} \hspace{3.3cm}  \psfig{figure=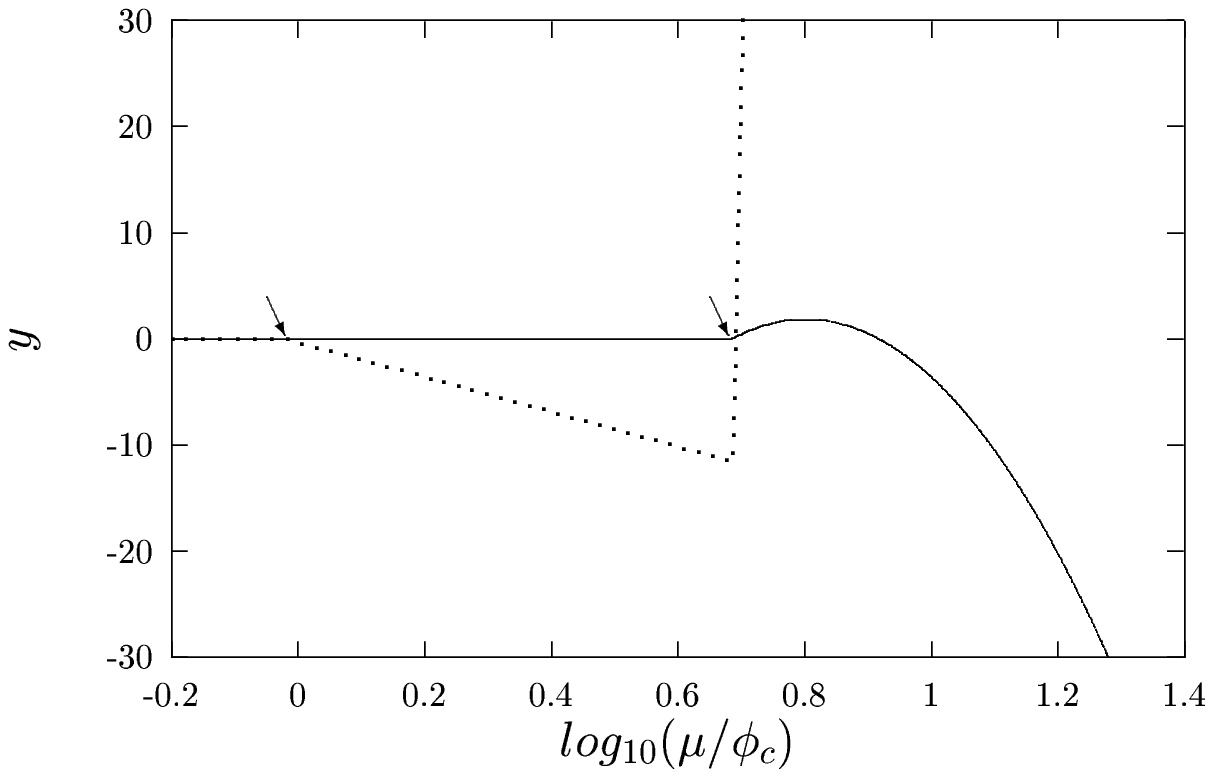,bbllx=5.6cm,bblly=16cm,bburx=13cm,bbury=24cm,height=6cm,width=4.5cm}}
\caption[]{\small{Plot of the total one-loop potential (solid line) and the 
tree level potential (dotted line) as a function of the scale $\mu$ for 
two different values of $\phi_c$. In the left panel
$\phi_c = 3\times 10^3$ GeV, so that $M_F>M_B$, while in the right panel
$\phi_c = 200$ GeV, so that $M_B>M_F$. The decoupling scales, 
corresponding to $M_B$ and $M_F$, are marked by arrows.
The vertical axis variable, $y$, represents the scalar potential
in a convenient choice of units as described in the text.}}
\label{fig:potential1}
\end{figure}

Let us now consider the one-loop effective potential, $V^{(0)}+V^{(1)}$,
for fixed $\phi_c$, as a function of $\mu$. The effective potential
undergoes the same decouplings as the $m,\lambda, g$ parameters. As
$\mu$ decreases, depending on the value of $\phi_c$, the fermion will be
decoupled first (the case $M_F>M_B$) or last (the case
$M_B>M_F$). Both cases are respectively illustrated~\footnote{The quantity
$y\equiv {\rm sign}(V-V(0))\times 
\log_{10}\left[\frac{|V-V(0)|}{TeV^4}+1\right]
10^{-6}$ that has been plotted in Fig.~\ref{fig:potential1}
has been chosen in order to give a continuous and faithful
representation of $V$ in logarithmic units.} in Fig.~\ref{fig:potential1},
where the solid lines denote the total one-loop potential
$V^{(0)}+V^{(1)}$ and the dotted lines the tree level approximation
$V^{(0)}$. The decoupling scales, corresponding to $M_B$ and $M_F$, are
marked by arrows. Incidentally, the one-loop potential between $M_B$ and
$M_F$ in Fig.~\ref{fig:potential1} (left panel)
 is a very slightly increasing function 
of $\mu$, which is
hardly appreciable from the plot. On the other hand, in 
Fig.~\ref{fig:potential1} (right panel) the
one-loop potential in the analogous region is exactly $\mu$-independent.
In either case, this is a peculiarity of the toy model at hand: in more
general models the one-loop potential is only $\mu$-flat in the $\mu\leq
M_B,M_F$ region, i.e. where all the particles get decoupled.

Let us now turn to the key question: which value of $\mu$, say $\mu^*$,
should be chosen to evaluate $V(\phi)$ in the most reliable way? As
explained in the previous section, there are two sensible criteria to
choose $\mu^*$:
\bea
&i)&\hspace{1cm}
V^{(1)}(\mu=\mu^*)=0
\nonumber\\
&ii)&\hspace{1cm}
\left.\frac{d\left[V^{(0)}+V^{(1)}\right]}{d\mu}\right|_{\mu=\mu^*}=0
\label{criteria}
\eea
They correspond respectively to the (apparently) best
perturbative convergence and the least $\mu$-sensitivity. Consistency of a
given choice of $\mu^*$ requires that both {\em i)} and {\em ii)} are
simultaneously satisfied, at least in an approximate way. 
From Fig.~\ref{fig:potential1}
it appears that there are two possible choices of $\mu^*$ satisfying
{\em i)} and {\em ii)} with various degrees of accuracy. First, as explained
in the previous section, any $\mu^*\leq M_B,M_F$ automatically fulfills 
{\em i)} and {\em ii)} in an exact way since all the particles get
decoupled for the radiative corrections. This is apparent from 
Fig.~\ref{fig:potential1}, since for   $\mu^*\leq M_B,M_F$ 
the solid and dotted lines coincide and are flat.
Second, there is another region of $\mu$, just above
$\max \{ M_B,M_F\}$,  where conditions {\em i)} and {\em ii)} are
approximately satisfied. This shows up in Fig.~\ref{fig:potential1}
as the region between the crossing of the solid line with the dotted line and
stationary point. Let us denote these two
regions as $(a)$ and $(b)$ respectively, i.e.

\begin{figure}[h]
\vspace{-3cm}
\mbox{
\psfig{figure=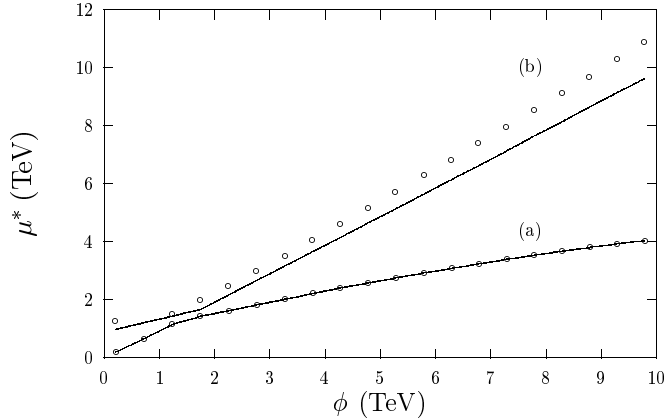,height=20cm}}
\vspace{-12cm}
\caption[]{\small{Plot of the $\mu^*$ as a function of $\phi$, according to the two choices (a) and (b) explained in the text.}}
\label{fig:muphi}
\end{figure}

Which region, $(a)$ or $(b)$, should be
selected to choose $\mu^*$? From Fig.~\ref{fig:potential1},
it could seem that this
question is not crucial since both choices provide similar values of
$V(\phi)$. However, in less simplified models this is not true
anymore. In fact, in models like the MSSM there may be not just two, but
several possible choices of $\mu^*$ approximately satisfying {\em i)}
and {\em ii)}, and leading to substantially different results. In order
to discriminate between $(a)$ and $(b)$ it is convenient to plot the
corresponding values of $\mu^*$ for different values of $\phi_c$, see
Fig.~\ref{fig:muphi}. For region $(a)$, condition {\em i)} (solid line) and 
condition {\em ii)} (circles) lead exactly to the same value of $\mu^*$,
namely any $\mu^*\leq M_B,M_F$. For the sake of definiteness we 
take $\mu^*=\min\{M_F,M_B\}$. On the contrary, for region $(b)$ [the region between solid line and circles]
conditions {\em i)} (solid line) and {\em ii)} (circles) lead to
different values of $\mu^*$. These values get increasingly split as
$\phi_c$ grows, eventually becoming enormously 
far away from each other. In other words,
in region $(b)$ criteria {\em i)} and {\em ii)} cannot be in general
simultaneously satisfied, which makes the evaluation of $V(\phi)$
unreliable. Significantly, as region $(b)$ becomes more and more
unreliable it also gets more and more separate from region $(a)$. 
Since $V(\phi)$ must be, of course, a continuous function of $\phi$,
region $(b)$ should be in general disregarded. It is worth-noticing that
the region of maximum reliability of region $(b)$, see Fig.~\ref{fig:muphi},
corresponds to the values of $\phi_c$ for which $M_B\sim M_F$. This is
satisfactory since in that case we do not have a multi-scale problem and
the usual choice $\mu^*\sim M_B,M_F$ is good enough.

In summary, we implement in the model the decoupling conditions at 
$M_B,M_F$, both for the $m,\lambda, g$ parameters and for the
one-loop effective potential. Then, we evaluate the potential at a scale
$\mu^*=\min\{M_F,M_B\}$, where all the particles get decoupled. Notice
that at this scale the tree-level and the one-loop potentials coincide
since $V^{(1)}=0$. So, at the end of the day we just need to evaluate
the tree-level potential $V^{(0)}$ at low energy in order to get the
best estimate of $V(\phi)$. In doing this it is important to keep in
mind that the values of the parameters at low energy
depend on the value of $\phi_c$ in the above-explained way, see
Figs.~\ref{fig:masa}--~\ref{fig:lambda}. 



\nsect{Relation with the multi-scale-renormalization approach}

The most sofisticated attempt up to now to deal with the effective
potential in the presence of several mass scales was introduced by
Einhorn and Jones (EJ)~\cite{EJ}, and recently improved by Ford and
Wiesendanger (FW)~\cite{FW1}. 
It is extremely interesting to study the relation of 
our method with theirs. In doing that we will
appreciate the underlying equivalence of both approaches and the
advantages of the decoupling approach proposed in this paper, 
both from the conceptual and the technical points of
view. This will also shed light on the physical significance of the
EJ and FW methods. 

The key point of the EJ and FW approaches is 
to extend the usual $\MSbar$ renormalization scheme by using several
{\it independent} renormalization
scales $\kappa_1, \kappa_2, ...$, 
which can then be adjusted differently to take
care of the different mass scales. 
Let us briefly review the EJ-FW method using the 
simple Higgs-Yukawa model with the lagrangian written in
Eq.~(\ref{lagr}). 

In the $\MSbar$ scheme the one-loop correction to
the effective potential is given by
\bea
V^{(1)}={\displaystyle 
\frac{1}{64\pi^2}\left[M_B^4\log\frac{M_B^2}{\mu^2} 
- 4N M_F^4\log\frac{M_F^2}{\mu^2}\right] }.
\label{unloopMS}
\eea
Here $\mu^2\equiv\bar\mu^2 e^{3/2}$, where $\bar\mu$ is the $\MSbar$
renormalization scale, and $M_B^2$, $M_F^2$ are the two mass scales 
in the theory, Eqs.~(\ref{MfMb}). Clearly, it is not possible to
track both scales in (\ref{unloopMS}) by a single choice of $\mu$.
The goal of the EJ-FW method in this case is to introduce two renormalization
scales, $\kappa_1, \kappa_2$, so that $V^{(1)}$ becomes
\bea
V^{(1)}={\displaystyle 
\frac{1}{64\pi^2}\left[M_B^4\log\frac{M_B^2}{\kappa_1^2} 
- 4N M_F^4\log\frac{M_F^2}{\kappa_2^2}\right] },
\label{unloopFW}
\eea
This would allow to take $\kappa_1=M_B$, $\kappa_2=M_F$, 
avoiding the undesirable presence of potentially large logarithms.
More precisely, the EJ idea is to modify the standard $\MSbar$ bare
lagrangian
\bea
{\cal L}_{\rm Bare}&=&
\frac{1}{2}Z_\phi(\partial\phi)^2 - \frac{1}{2}Z_\phi Z_{m^2}m^2\phi^2 -
\frac{1}{4!}\bar\mu^{\epsilon}Z_\phi^2Z_{\lambda}\lambda\phi^4 
\nonumber \\
&+& 
Z_\psi\bar\psi i\slash\!\!\!\partial \psi +
Z_\psi Z_gZ_\phi^{\frac{1}{2}} \bar\mu^{\frac{1}{2}\epsilon} g\bar\psi
\phi\psi - \Lambda + (Z_\Lambda-1)\bar\mu^{-\epsilon}m^4\lambda^{-1}
\ ,
\label{lagrBare}
\eea
replacing the occurrences of $\bar\mu$ with independent scales
$\kappa_1, \kappa_2$, ..., in such a way that the one-loop effective
potential gets the form (\ref{unloopFW}). Actually, they got something
similar to Eq.~(\ref{unloopFW}) but with extra terms involving the
potentially large $\log\kappa_1^2/\kappa_2^2$. Hence,
their attempt was not completely succesful. FW modified the EJ scheme
allowing the RG independent scales $\kappa_1, \kappa_2$ to be attached
to the kinetic terms in a clever way. In particular they proposed to
re-write ${\cal L}_{\rm Bare}$ as
\bea
{\cal L}_{\rm Bare}^{\small{\rm FW}}&=&
\frac{1}{2} \kappa_1^{-\frac{1}{2}\epsilon} Z_\phi(\partial\phi)^2 -
\frac{1}{2}Z_\phi Z_{m^2}m^2\phi^2 -
\frac{1}{4!}Z_\phi^2Z_{\lambda}\lambda\phi^4 
\nonumber \\
&+& 
\kappa_2^{-\frac{1}{4}\epsilon}Z_\psi\bar\psi i\slash\!\!\!\partial \psi +
Z_\psi Z_gZ_\phi^{\frac{1}{2}} g\bar\psi
\phi\psi - \Lambda - (Z_\Lambda-1)m^4\lambda^{-1}
\ .
\label{lagrBareFW}
\eea
The relation between the bare and renormalized parameters is then given
by 
\bea
&\lambda_B&=\kappa_1^{\epsilon}Z_{\lambda}\lambda,\;\;\;
g_B^2=\kappa_1^{\frac{1}{2}\epsilon}\kappa_2^{\frac{1}{2}\epsilon}
Z_{g^2} g^2,
\nonumber \\
&m_B^2&=\kappa_1^{\frac{1}{2}\epsilon}Z_{m^2}m^2,\;\;\;
\phi_B=\kappa_1^{-\frac{1}{4}\epsilon}
Z_{\phi}^{\frac{1}{2}} \phi,
\nonumber \\
&\psi_B&=\kappa_2^{-\frac{1}{8}\epsilon}
Z_{\psi}^{\frac{1}{2}} \psi,\;\;\;
\Lambda_B=\Lambda + (Z_\Lambda-1)m^4\lambda^{-1}
\ .
\label{bareren}
\eea
Then, the one-loop effective potential really gets the form
(\ref{unloopFW}). The use of two independent RG scales $\kappa_1,
\kappa_2$ implies double number of RGEs.
Namely, for each generic parameter $h$ there are two $\beta$-functions,
defined as
\be
{}_i\beta_h^{\small{\rm FW}}= \kappa_i\frac{d}{d\kappa_i}h,\;\;\;\;i=1,2
\label{betah}
\ee
Consistency of the approach requires ${}_1\beta_h+{}_2\beta_h=\beta_h$,
i.e. the complete $\MSbar$ $\beta$-function. For the
particular choice $\kappa_1=\kappa_2$ the standard $\MSbar$ scheme 
is recovered.
The ${}_i\beta$ functions are evaluated from the $Z$ factors in a
standard way. For example,
\bea
{}_1\beta_\lambda^{\small{\rm FW}} &=& 
-\epsilon\lambda-\lambda\kappa_1\left[\frac{\partial z_\lambda}{\partial
\lambda}\frac{\partial \lambda}{\partial \kappa_1} +
\frac{\partial z_\lambda}{\partial
g^2}\frac{\partial g^2 }{\partial \kappa_1} + ... \right] =
-\epsilon\lambda
+\lambda(\lambda\partial_\lambda
+\frac{1}{2}g^2\partial_{g^2})z_\lambda\;,
\nonumber \\
{}_2\beta_\lambda^{\small{\rm FW}} &=& 
\lambda\kappa_2\left[\frac{\partial z_\lambda}{\partial
\lambda}\frac{\partial \lambda}{\partial \kappa_2} +
\frac{\partial z_\lambda}{\partial
g^2}\frac{\partial g^2 }{\partial \kappa_2} + ... \right] =
\frac{1}{2}\lambda 
g^2\partial_{g^2}z_\lambda,
\label{betalambdag}
\eea
where $z_\lambda$ denotes the $1/\epsilon$ coefficient in $Z_\lambda$. 
The one-loop contribution to the $z$'s is identical to the $\MSbar$
one. In particular
\be
z_\lambda=\frac{1}{(4\pi)^2}(3\lambda+8N g^2-48N g^4 \lambda^{-1})
\label{zetalambda}
\ee
Thus,
\be
{}_1\beta_\lambda^{\small{\rm FW}} = 
\frac{1}{(4\pi)^2}(3\lambda^2+4N g^2\lambda)
,\;\;\;
{}_2\beta_\lambda^{\small{\rm FW}} = \frac{1}{(4\pi)^2}(4Ng^2\lambda-48N g^4)
\label{betalambda}
\ee
For the other parameters the expressions are analogous (for more
details, see Ref.~\cite{FW1}).

FW give an additional condition for the consistency of their approach,
namely
\bea
\left[{\cal D}_1,{\cal D}_2\right]=0
\label{conm}
\eea
where ${\cal D}_i\equiv d/d \kappa_i$. This arises as an integrability
condition for the RGE satisfied by the effective potential, ${\cal D}_i
V=0$. As was shown by FW, the implementation of Eq.~(\ref{conm}) requires
in general that beyond one-loop-order the ${}_i\beta$-functions contain
logarithms of the $\kappa_i$ ratios. We will turn to this point 
in section~5.

The previous FW procedure is ingenious and successful, in the sense that
it leads to the one-loop effective potential written in the form
(\ref{unloopFW}). However, it is clearly involved and seems hard to 
be extended to more complicated models, such as the SM and the
MSSM. Moreover, there is a lack of intuition about the physical role of
the several $\kappa_i$ scales. In this sense, attempts to associate the
$\kappa_i$ running with the usual $\MSbar$ running when all the
particles but one (the one whose mass is tracked by $\kappa_i$) are
decoupled, were unfruitful~\cite{FW2}. For example, in the case at hand the
${}_i\beta_\lambda$-functions given in Eq.~(\ref{betalambda}) 
do not correspond to
the presence of just the boson or the fermion respectively. If they did,
the two beta functions would have been $(4\pi)^2{}_1\beta_\lambda\rightarrow
3\lambda^2$, $(4\pi)^2{}_2\beta_\lambda\rightarrow 
8g^2N(\lambda-6g^2)$ [see Eq.~(\ref{RGE})]. This shortcoming prevents from
giving a clear physical interpretation to the $\kappa_i$ scales. We will
come back to this point later.

Let us now turn to the question: Is it possible to reproduce the results
of the EJ-FW multi-scale renormalization with our decoupling
approach? We will see that the answer is yes. First of all, let
us see how our decoupling approach can be interpreted as a scenario with
several {\em independent} scales. The key point is to allow the
decoupling scales to be free parameters. Let us illustrate this by
using the simple Higgs-Yukawa model. Then, instead of decoupling the 
boson (fermion) at $\mu=M_B$ ($\mu=M_F$), as described in section~3, we
will do it at $\mu=\mu_B$ ($\mu=\mu_F$), where $\mu_B$, $\mu_F$ are in
principle arbitrary. This procedure is acceptable since in a
mass-independent renormalization scheme, such as the $\MSbar$, 
the decoupling is implemented by hand and there is no indication about
the value of the decoupling scale. Besides, the complete effective
potential is $\mu$-independent, which in principle allows to decouple at
any scale.

Decoupling at scales $\mu_B$, $\mu_F$ for the one-loop effective
potential amounts to replace the 
$\theta_{B,F}=\theta(\mu-M_{B,F})$ functions in (\ref{unloopmod}) by
\bea
\theta_{B,F}=\theta(\mu-\mu_{B,F})
\label{thetas}
\eea
Then, there appear threshold
corrections in $V^{(1)}$ [see Eq.~(\ref{unloopmod})]. Keeping these
threshold pieces explicitly, the one-loop effective potential,
$V^{(0)}+V^{(1)}$, reads
\bea
V^{(0)}&=&\frac{1}{2}m^2\phi^2+\frac{1}{4!}\lambda\phi^4+\Lambda
\nonumber\\
V^{(1)}&=& 
\frac{1}{64\pi^2}\left[M_B^4\log\frac{M_B^2}{\mu_B^2} +
M_B^4\log\frac{\mu_B^2}{\mu^2}\theta_B \right.
\nonumber\\
&-& \left. 4N M_F^4\log\frac{M_F^2}{\mu_F^2}
- 4N M_F^4\log\frac{\mu_F^2}{\mu^2}\theta_F\right] \ .
\label{unloopmod2}
\eea
The one-loop  $\beta$-functions are extracted from (\ref{unloopmod2}) in
the usual way, applying the RG condition $dV/d\mu=0$. They are exactly
as in Eq.~(\ref{RGE}), but with $\theta_{B,F}$ defined as in
(\ref{thetas}). Notice that all of them have the same structure. Namely,
for a generic coupling $h$, the $\beta$-function has the form
\bea
\beta_{h}=\mu\frac{dh}{d\mu}={}_B\beta_{h}\theta_B + {}_F\beta_{h}\theta_F. 
\label{betahBF}
\eea
At low scale, i.e. $\mu_m\leq\mu_B, \mu_F$, $V^{(1)}$ gets the form
\bea
V^{(1)}={\displaystyle 
\frac{1}{64\pi^2}\left[M_B^4\log\frac{M_B^2}{\mu_B^2} 
- 4N M_F^4\log\frac{M_F^2}{\mu_F^2}\right] },
\label{unloopBF}
\eea
This is exactly as Eq.~(\ref{unloopFW}) from the FW approach, where the
role of $\kappa_{1,2}$ is now played by $\mu_{B,F}$. Since $\mu_{B,F}$
are arbitrary scales we can choose $\mu_{B}=M_B$, $\mu_{F}=M_F$, getting
rid of the potentially large logarithms and recovering the decoupling
approach in the form described in sections 2, 3~\footnote{Incidentally,
this shows that the $M_B$, $M_F$ scales are the most convenient ones
to implement the decoupling from the calculational point of view. This
point was mentioned in section~2.}. Hence, at this stage we
notice a perfect equivalence between the FW approach and our method.
Concerning the ${}_i\beta$-functions, defined in Eq.~(\ref{betah}) in the
FW approach, we must study here the dependence of the various parameters
with respect to $\mu_{B}$, $\mu_{F}$. Notice that a generic parameter
$h(\mu=\mu_m)$ is given by
\bea
h(\mu_m)=h(\mu_M)-\int_{\mu_m}^{\mu_M}d\mu\frac{dh}{d\mu}\;\;,
\label{hlow}
\eea
where $\mu_M$ is the initial high scale. So, using $d
\theta(\mu-\mu_{B,F})/d\mu_{B,F}=\delta(\mu-\mu_{B,F})$, it is
clear that
\bea
\mu_B\frac{dh}{d\mu_B}&=&{}_B\beta_{h}
\nonumber \\
\mu_F\frac{dh}{d\mu_F}&=&{}_F\beta_{h}
\label{betahBF2}
\eea
where ${}_{B,F}\beta_{h}$ have been defined in Eqs.~(\ref{betahBF}),
(\ref{RGE}).
Obviously, ${}_{B}\beta_{h}+{}_{F}\beta_{h}=\beta_{h}$, again as in the
FW approach. However, there is an important difference: the values of
${}_{B,F}\beta$ for the various parameters of the model do {\em not}
coincide with those of ${}_{1,2}\beta^{\small{\rm FW}}$ in the FW approach
[see Eq.~(\ref{betalambda})]. For example,
\be
{}_B\beta_\lambda = \frac{1}{(4\pi)^2} 3\lambda^2
,\;\;\;
{}_F\beta_\lambda = \frac{1}{(4\pi)^2}8g^2N(\lambda-6g^2)
\label{betalambdaBF}
\ee
[to be compared to (\ref{betalambda})].
It is interesting to note that in our approach the
${}_{B}\beta$- and ${}_{F}\beta$-functions do correspond to the
$\beta$-functions when all the particles except
the boson or the fermion (respectively) are 
decoupled. This is precisely what was (unsuccessfully)
required in Ref.~\cite{FW2} to interpret the EJ-FW method as an implementation
of the decoupling. In our case this emerges automatically by
construction.

Our final goal in this section is to prove that there is a (slight
variation of the) FW approach that is exactly equivalent to our
decoupling scheme. First of all, notice that the FW lagrangian of
Eq.~(\ref{lagrBareFW}) is not the only consistent
multiscale-renormalization proposal. For
example, one may simply replace $\kappa_1^{-\frac{1}{2}\epsilon}
\rightarrow \kappa_1^{-(\frac{1}{2}+x)\epsilon}\kappa_2^{-x\epsilon}$,
$\kappa_2^{-\frac{1}{4}\epsilon}
\rightarrow \kappa_1^{-(\frac{1}{4}+y)\epsilon}\kappa_2^{-y\epsilon}$
in Eq.~(\ref{lagrBareFW}), which leads to straightforward modifications
in Eq.~(\ref{bareren}). In this case, it is easy to verify that $x=0$,
$y=0$ are in fact the only constant values of $x,y$ leading to the desired form
(\ref{unloopFW}) for $V^{(1)}$. 
%
%
However, there is still a
possible modification of the lagrangian of Eq.~(\ref{lagrBareFW}) that
apparently was not considered by FW. Namely, let us replace
(\ref{lagrBareFW}) by
\bea
{\cal L}_{\rm Bare}&=&
\frac{1}{2} \kappa_1^{-\frac{1}{2}\epsilon}f_1(\kappa_1,\kappa_2) 
Z_\phi(\partial\phi)^2 -
\frac{1}{2}Z_\phi Z_{m^2}m^2\phi^2 -
\frac{1}{4!}Z_\phi^2Z_{\lambda}\lambda\phi^4 
\nonumber \\
&+& 
\kappa_2^{-\frac{1}{4}\epsilon}f_2(\kappa_1,\kappa_2)
Z_\psi\bar\psi i\slash\!\!\!\partial \psi +
Z_\psi Z_gZ_\phi^{\frac{1}{2}} g\bar\psi
\phi\psi - \Lambda - (Z_\Lambda-1)m^4\lambda^{-1}
\ ,
\label{lagrBareFW2}
\eea
where $f_1,f_2$ are dimensionless functions of $\kappa_1$, $\kappa_2$.
They must satisfy $f_{1,2}(\kappa,\kappa)=1$ in order 
to ensure that the $\MSbar$ scheme is recovered for
$\kappa_1=\kappa_2$. Moreover, the usual tree-level lagrangian must be
recovered when the one-loop corrections are neglected. This means that
the $f$-functions can only be non-trivial beyond tree-level; in other
words, $f_{1,2}=1\;+\;{\cal O}(\hbar)$.

It is now straightforward to show that taking
\bea
f_{1}(\kappa_1,\kappa_2)&=& \left[\frac{\kappa_2}{\kappa_1}\right]^{a/2}
\nonumber\\
f_{2}(\kappa_1,\kappa_2)&=& \left[\frac{\kappa_1}{\kappa_2}\right]^{a/4}
\frac{\kappa_1^{b/2}\kappa_2^{c/2}}{[\kappa_1\theta_{12} +
\kappa_2\theta_{21}]^{(b+c)/2}} 
\label{f1f2}
\eea
where
\bea
a&=&\frac{1}{(4\pi)^2}4N g^2
\nonumber\\
b&=&\frac{1}{(4\pi)^2}(-2N-3)g^2 
\nonumber\\
c&=&\frac{1}{(4\pi)^2}(2N-3)g^2 
\nonumber\\
\theta_{ij}&=&\theta(\kappa_i-\kappa_j)
\label{notations}
\eea
the ${}_{1}\beta$- and ${}_{2}\beta$-functions for all the parameters
exactly coincide with the ${}_{B}\beta$- and ${}_{F}\beta$-functions
readable from Eq.~(\ref{RGE}). To check that, notice that from 
(\ref{lagrBareFW2}), (\ref{f1f2}) the relation between the bare and
renormalized parameters is now given by [compare to (\ref{bareren2})]
\bea
&\lambda_B&=\kappa_1^{\epsilon}\left[\frac{\kappa_2}{\kappa_1}\right]^{-a}
Z_{\lambda}\lambda,\;\;\;
g_B^2=\kappa_1^{\frac{1}{2}\epsilon}\kappa_2^{\frac{1}{2}\epsilon}
\frac{[\kappa_1\theta_{12} +
\kappa_2\theta_{21}]^{b+c}}{\kappa_1^{b}\kappa_2^{c}}
Z_{g^2} g^2,
\nonumber \\
&m_B^2&=\kappa_1^{\frac{1}{2}\epsilon}
\left[\frac{\kappa_2}{\kappa_1}\right]^{-a/2}
Z_{m^2}m^2,\;\;\;
\phi_B=\kappa_1^{-\frac{1}{4}\epsilon}
\left[\frac{\kappa_2}{\kappa_1}\right]^{a/4}
Z_{\phi}^{\frac{1}{2}} \phi,
\nonumber \\
&\psi_B&=\kappa_2^{-\frac{1}{8}\epsilon}
\left[\frac{\kappa_1}{\kappa_2}\right]^{a/8}
\frac{\kappa_1^{b/4}\kappa_2^{c/4}}{[\kappa_1\theta_{12} +
\kappa_2\theta_{21}]^{(b+c)/4}}
Z_{\psi}^{\frac{1}{2}} \psi,\nonumber\\ & & \nonumber\\
&\Lambda_B&=\Lambda + (Z_\Lambda-1)m^4\lambda^{-1}
\ .
\label{bareren2}
\eea
Then, the $\beta$-functions are modified in a straightforward way. For
example,
\bea
{}_1\beta_\lambda&=&{}_1\beta_\lambda^{\small{\rm FW}}
-\frac{1}{(4\pi)^2} 4g^2\lambda N= 
\frac{1}{(4\pi)^2}3\lambda^2 
\nonumber\\
{}_2\beta_\lambda&=&{}_2\beta_\lambda^{\small{\rm FW}}+
\frac{1}{(4\pi)^2}4 g^2\lambda N =
\frac{1}{(4\pi)^2} 8 g^2 N \left(\lambda-6 g^2 \right) 
\nonumber\\
{}_1\beta_{m^2}&=&{}_1\beta_{m^2}^{\small{\rm FW}}
-\frac{1}{(4\pi)^2}2 g^2 N m^2 =
\frac{1}{(4\pi)^2}  \lambda m^2
\nonumber \\
{}_2\beta_{m^2}&=&{}_2\beta_{m^2}^{\small{\rm FW}}+
\frac{1}{(4\pi)^2}2g^2 Nm^2 =
\frac{1}{(4\pi)^2} 4 N g^2 m^2
\nonumber\\
{}_1\gamma_{\phi}&=&{}_1\gamma_{\phi}^{\small{\rm FW}}
-\frac{1}{(4\pi)^2}g^2 N =0
\nonumber \\
{}_2\gamma_{\phi}&=&{}_2\gamma_{\phi}^{\small{\rm FW}}+
\frac{1}{(4\pi)^2}g^2 N =
\frac{1}{(4\pi)^2} 2  g^2 N
\label{RGE2}
\eea
Thus, lagrangian (\ref{lagrBareFW2}) with $f_{1,2}$ defined as in
Eq.~(\ref{f1f2}) exactly  implements the decoupling approach.
Still, the one-loop potential $V^{(1)}$ maintains the desired form
(\ref{unloopFW}). Of course the renormalization scheme defined by
Eqs.~(\ref{lagrBareFW2}), (\ref{f1f2}) may a priori seem absolutely ad
hoc. Its usefulness is to prove the equivalence of our decoupling
approach (which does emerge in a natural way) with a particular
multi-scale renormalization scheme.

\nsect{Two-loop effects}

The procedure of section 3 can be straightforwardly generalized
to include two-loop corrections to the effective potential. In
particular, in the Higgs-Yukawa model, these corrections have
been computed in Ref.~\cite{FIJ} as
\be
V^{(2)}=\kappa^2 \left(\Vbd + \Vbf \right)
\label{dosloops}
\ee
where $\Vbd$ includes the sunset and figure eight diagrams with
boson internal lines, and $\Vbf$ is the sunset diagram with a
fermion loop line. The precise expressions for these terms can
be found in Ref.~\cite{FIJ}. Using the fact that the decoupling
of $\Vbd$ is controlled by the scale $M_B$ while that of $\Vbf$
is controlled by both $M_B$ and $M_F$, we can generalize the
one-loop improved potential in the presence of decoupling, given
by Eqs.~(\ref{unloopdec}), (\ref{unloopmod}), as:
\bea
&V^{(1)}+V^{(2)}&=\kappa\left[\Vbu + \kappa\Vbd
+\kappa\theta(M_B-M_F) \Vbf \right]\;\theta_B^{(2)}\nonumber\\
&&+\kappa\left[\Vfu +\kappa\theta(M_F-M_B)\Vbf \right]\;
\theta_F^{(2)}
\label{total2}
\eea
where the $\theta$-functions are now defined by,
\bea
\theta_B^{(2)}&=&\theta\left( -\Vbu -\kappa\Vbd
-\kappa\theta(M_B-M_F) \Vbf \right) \nonumber\\
\theta_F^{(2)}&=&\theta\left( -\Vfu-\kappa\theta(M_F-M_B) \Vbf
\right)\ . 
\label{thetas2}
\eea
From the property $\theta\left(f(\mu)\right)=\theta(\mu-\mu_0)$,
where $f(\mu)$ is a monotonically increasing function with
$f(\mu_0)=0$, we can write (\ref{thetas2}) in a two-loop
expansion as $\theta_B^{(2)}=\theta\left(\mu-\mu_B^{(2)}\right)$,
$\theta_F^{(2)}=\theta\left(\mu-\mu_F^{(2)}\right)$, 
where now the decoupling scales $\mu_{B,F}^{(2)}$ are given by the
two-loop expansion
\bea
\mu_B^{(2)}&=& M_B+\left.\frac{2}{M_B^3} \kappa \left[
\Vbd + \theta(M_B-M_F) \Vbf \right] \right|_{\mu=M_B}+{\cal
O}({\hbar}^2) \nonumber\\
\mu_F^{(2)}&=& M_F-\left.\frac{1}{2NM_F^3} \kappa 
\theta(M_F-M_B) \Vbf  \right|_{\mu=M_F}+{\cal
O}({\hbar}^2)
\label{scales2l}
\eea
Decoupling bosons and fermions at scales (\ref{scales2l}) there
are no threshold corrections at two-loop, neither 
in the effective potential nor in the coupling constants, since
the property $f {\cal D}\theta (f)=f\delta(f){\cal D}f\equiv 0$
guarantees the absence of $\delta$-functions in the RGE.
It might seem that by taking the decoupling conditions (\ref{thetas2}) 
we can get rid of the potentially large logarithms at all.
This would contradict the statement of Ref.~\cite{FW1} concerning
the unavoidable appearance of logarithms beyond one-loop order.
However, there is no contradiction, since the logarithms are now
hidden in the definitions of $\mu_B^{(2)}$ and $\mu_F^{(2)}$.
Nevertheless, it is amazing to check the logarithms do never spoil
the perturbative validity of the approach, since they cancel each other
in the potentially dangerous limits $M_F/M_B \to 0,\infty$. To see
this, it is convenient to explicitly write the decoupling scales
(\ref{scales2l})
\bea
\mu_B^{(2)}&=&M_B\left[ 1+\frac{4\kappa g^2 N \theta(M_B-M_F)}{M_B^4}
 \left\{(3M_F^4 - M_B^2 M_F^2)\log \frac{M_F^2}{M_B^2} \right. \right. 
\nonumber \\
&-& \left. \left. (M_F^4 - \frac{3}{2} M_B^2 M_F^2 + \frac{1}{4} M_B^4)
\log^2\frac{M_F^2}{M_B^2} 
- \frac{1}{2} (2M_F^2-\frac{1}{2}M_B^2) \xi(M_F,M_B)\right\} \right.  
\nonumber \\  
&+& \left. {\rm non-logarithms} \phantom{\frac{1}{2}} \!\!\!\! \right] 
\nonumber \\
\mu_F^{(2)}&=&M_F\left[ 1-\frac{\kappa g^2\theta(M_F-M_B)}{2 M_F^4}
\left\{ (M_B^2 M_F^2- \frac{1}{2} M_B^4)\log\frac{M_B^2}{M_F^2} \right. \right.
\nonumber \\
&-& \left.\left. \frac{1}{2}(2M_F^2-\frac{1}{2}M_B^2) \xi(M_F,M_B) 
\right \}+{\rm non-logarithms}\right],
\label{scales3}
\eea
where $\xi(M_F,M_B)$ is defined as
\bea
\xi(M_F,M_B) = 4{\left|M_B^4-4M_F^2 M_B^2 \right|}^{1/2} \!\! \left \{ 
\begin{array}{ll} \!\!\!
-M(-\vartheta_F) + M(\vartheta_F) + M(\vartheta_B) &  
\!\! \textrm{if $M_B^2 > 4M_F^4$} \\
\!\!\! 2L(\varphi_F) + L(\varphi_B) - \frac{\pi}{2} \log 2 & \! \! 
\textrm{otherwise}
\end{array} \right.
\label{horrible}
\eea 
with
\bea
M(t) &=& -\int_{0}^{t}{dx\;\log \sinh x}, \qquad \qquad  
L(t) = -\int_{0}^{t}{dx\;\log \cos x}, \nonumber\\
\vartheta_F &=& \coth^{-1}\left(\frac{M_B^2}{{(M_B^4-4M_F^2 M_B^2)}^{1/2}}
\right), \qquad 
\vartheta_B = \coth^{-1}\left(\frac{2M_F^2-M_B^2}{{(M_B^4-4M_F^2 M_B^2)}^{1/2}}
\right), \phantom{2222} \nonumber\\
\varphi_F &=& \tan^{-1}\left(\frac{M_B^2}{{(4M_F^2 M_B^2 - M_B^4)}^{1/2}}
\right), \qquad 
\varphi_B = \tan^{-1}\left(\frac{2M_F^2-M_B^2}{{(4M_F^2 M_B^2 - M_B^4)}^{1/2}}
\right).
\label{angulos}
\eea
It is now straightforward to check that in the limit $M_F/M_B \to 0$
\bea
\xi(M_F,M_B) \to M_B^2 \left(\frac{\pi^2}{3} + \log^2 \frac{M_F^2}{M_B^2}
+{\cal O}\left(\frac{M_F^2}{M_B^2} \right ) \right ),
\label{limite}
\eea
so that
\bea
\mu_B^{(2)} &\to& M_B \left [ 1 + 
\kappa \left \{ {\rm constant} + {\cal O}\left(
\frac{M_F^2}{M_B^2} \right ) \right \} \right ]
\label{limscale1} \nonumber \\
\mu_F^{(2)} &\to& M_F
\eea
Analogously, in the other potentially dangerous case, $M_F/M_B \to \infty$,
one gets
\bea
\xi(M_F,M_B) \to 0
\label{limite2}
\eea
and
\bea
\mu_F^{(2)} &\to& M_F \left [ 1 + \kappa \left \{ {\rm constant} + 
{\cal O}\left(
\frac{M_B^2}{M_F^2} \right ) \right \} \right ]
\label{limscale2} \nonumber \\
\mu_B^{(2)} &\to& M_B \left[1 + \kappa \left \{{\rm constant}\right \} \right]  
\eea 
Eqs.~(\ref{limscale1}), (\ref{limscale2}) show that the potentially 
dangerous limits $M_F/M_B \to 0,\infty$ are in fact well-behaved.
Consequently, our decoupling approach remains consistent with perturbation
theory, even when the relevant scales are very different. This is
highly satisfactory.

It is interesting to mention that had we kept the one-loop choice
of decoupling scales for the two-loop level, 
i.e. $\mu_{B,F}^{(2)} = M_B,M_F$,  we would have found
non-vanishing threshold effects involving $\log M_F^2/M_B^2$ terms.
These terms are in direct correspondence to the logarithms found in
the two-loop ${}_{i}\beta$ - functions of the FW approach. 
They are in fact needed to implement the FW condition (\ref{conm}).
On the other hand, the $\mu_{B,F}^{(2)}$ scales defined in
(\ref{scales2l}) are not bound to satisfy condition (\ref{conm}),
which is consistent with the absence of logarithmic threshold corrections.

\nsect{Conclusions}

In this paper we have dealt with the problem of multi-scale theories in
mass-independent renormalization schemes. We have proposed a method based on
the use of the decoupling theorem at the field-dependent scales
$M_i(\phi_c)$, which are the relevant mass-scales in effective potential
calculations. The successive decouplings of the fields lead to a simple
prescription for the scale at which the effective potential has the least
scale-dependence and the loop expansion has the best behaviour, as
min$_i\{M_i(\phi_c)\}$. Our method does not yield any threshold
corrections for any of the parameters appearing in the effective
potential, a feature wich can be implemented to any arbitrary
order in perturbation theory.

We have compared our decoupling approach with previously proposed methods
to deal with the multi-scale problem in effective potential calculations.
In particular we have compared our approach with the multi-scale
renormalization approach and found the particular multi-scale formalism
equivalent to our decoupling approach at one-loop. However, unlike the
multi-scale formalism, our decoupling method can avoid the presence of
large logarithms corresponding to very different scales, as we have
explicitly exhibited at two-loop in the Higgs-Yukawa model. On the
other hand previous methods to deal with the multi-scale problem based 
on decoupling used as decoupling scales invariant masses in the lagrangian. 
Clearly these methods were insufficient to extend the effective potential
to large values of the field, where field-dependent masses are larger than
field independent ones, or to decouple any particle without an invariant
mass.

We have chosen a particularly simple case, the Higgs-Yukawa model,
to illustrate the main features of the method, but clearly
it can be implemented in more realistic models as
the Standard Model and the MSSM. In both cases the method present in some
cases advantages over previous studies. In the case of the SM, as we have
stated above, our method allows to go to large values of the classical   
Higgs field while properly decoupling all particles at their
corresponding mass scales. In particular the problem of the stability of
the electrowek minimum and its relation to the scale of new physics can
be safely considered. A result which can be of great interest if a light
Higgs is found at present or future colliders. In the case of the MSSM
our method is appropriate to deal with some extreme situations in  
the calculations of the radiatively corrected Higgs mass spectrum,
in particular with those involving a large mixing in the left-right    
stop sector. A case which is difficult to deal with in present
effective potential calculations. Those issues will be the object of
future investigations.

 

\end{document}